\documentclass[%
 reprint,
 superscriptaddress,
 amsmath,amssymb,
 aps,
floatfix,
]{revtex4-2}

\usepackage{graphicx}   
\usepackage{dcolumn}    
\usepackage{bm}         

\usepackage[table,dvipsnames]{xcolor}
\usepackage{booktabs}
\usepackage{siunitx}
\usepackage{hyperref}
\usepackage{amsmath}
\usepackage[capitalize]{cleveref}
\usepackage{float}
\usepackage{amssymb}
\usepackage{threeparttable}

\usepackage[nolist]{acronym}

\usepackage{mathtools}

\newcommand{\action}{\bm{a}}
\newcommand{\actuators}{\bm{u}}
\newcommand{\beam}{\bm{b}}
\newcommand{\beamheight}{\sigma_y}
\newcommand{\beampositionx}{\mu_x}
\newcommand{\beampositiony}{\mu_y}
\newcommand{\beamwidth}{\sigma_x}

\newcommand{\kick}{\alpha}
\newcommand{\incomingbeam}{\beam_\textit{in}}
\newcommand{\measuredbeam}{\beam}
\newcommand{\misalignments}{\bm{m}}
\newcommand{\nninput}{\bm{x}}
\newcommand{\nnoutput}{\bm{y}}
\newcommand{\normalizationscalevector}{\bm{\lambda}}
\newcommand{\normalizedactuators}{\actuators_\text{normed}}
\newcommand{\observation}{\bm{o}}
\newcommand{\outgoingbeam}{\beam_\text{out}}

\newcommand{\quadrupole}{Q}
\newcommand{\reward}{r}
\newcommand{\spacechargedelta}{\Delta\Sigma_\text{SC}}
\newcommand{\state}{\bm{s}}
\newcommand{\steerer}{C}

\newcommand{\targetbeam}{\beam'}
\newcommand{\transfermap}{R}
\newcommand{\quadstrength}{k}
\newcommand{\weights}{\bm{w}}

\begin{document}

\preprint{APS/123-QED}

\title{Cheetah: Bridging the Gap Between Machine Learning and Particle Accelerator Physics with High-Speed, Differentiable Simulations}
\thanks{All figures and pictures by the authors are published under a CC-BY7 licence.}

\author{Jan Kaiser}
\email{jan.kaiser@desy.de}
\thanks{Equal contributions}
\affiliation{Deutsches Elektronen-Synchrotron DESY, Germany}

\author{Chenran Xu}
\email{chenran.xu@kit.edu}
\thanks{Equal contributions}
\affiliation{Karlsruhe Institute of Technology (KIT), Germany}

\author{Annika Eichler}
\affiliation{Deutsches Elektronen-Synchrotron DESY, Germany}
\affiliation{Hamburg University of Technology, 21073 Hamburg, Germany}

\author{Andrea \surname{Santamaria Garcia}}
\affiliation{Karlsruhe Institute of Technology (KIT), Germany}

\date{11 January 2024}

\begin{abstract}
    Machine learning has emerged as a powerful solution to the modern challenges in accelerator physics.
    However, the limited availability of beam time, the computational cost of simulations, and the high-dimensionality of optimisation problems pose significant challenges in generating the required data for training state-of-the-art machine learning models.
    In this work, we introduce \textit{Cheetah}, a PyTorch-based high-speed differentiable linear-beam dynamics code.
    Cheetah enables the fast collection of large data sets by reducing computation times by multiple orders of magnitude and facilitates efficient gradient-based optimisation for accelerator tuning and system identification.
    This positions Cheetah as a user-friendly, readily extensible tool that integrates seamlessly with widely adopted machine learning tools.
    We showcase the utility of Cheetah through five examples, including reinforcement learning training, gradient-based beamline tuning, gradient-based system identification, physics-informed Bayesian optimisation priors, and modular neural network surrogate modelling of space charge effects.
    The use of such a high-speed differentiable simulation code will simplify the development of machine learning-based methods for particle accelerators and fast-track their integration into everyday operations of accelerator facilities.
\end{abstract}

\maketitle


\section{Introduction}\label{sec:introduction}

Future particle accelerator experiments will place ever-increasing demands on the performance and capabilities of particle accelerator operations and experiment analysis.
In order to meet these demands, the research community is increasingly turning to \ac{ML} methods, which have already demonstrated their ability to push the envelope of what is possible in the field of accelerator science~\cite{kaiser2022learningbased,edelen2017using,fujita2021physicsinformed,kaiser2023machine}.

One of the remaining challenges holding back this line of research is the demand for large amounts of data (including environment interactions) of these methods.
\Acf{RL}, for example, has already successfully been used to train intelligent tuning algorithms and controllers that can outperform the currently deployed black-box optimisation algorithms and handcrafted controllers~\cite{kaiser2022learningbased,kaiser2023learning,xu2023beam,kain2020sampleefficient}.
However, \ac{RL} methods require many interactions with their target task to train a well-performing policy.
For example, \num{6000000} samples were needed in~\cite{kaiser2022learningbased} to successfully train a policy on a transverse beam tuning task.
Orders of magnitude larger number of samples are also common with other \ac{RL} applications~\cite{degrave2022magnetic,vinyals2019grandmaster}.
The general scarcity of beam time makes collecting experimental data for \ac{ML} methods, such as \ac{RL}, a significant bottleneck.
Gathering at least partial data sets in simulation can alleviate this problem, but existing accelerator simulation codes have mostly been developed with a focus on the design phase of accelerators, where high-fidelity and physical correctness are critical, and computing times range from minutes to several hours for one simulation.
Consequently, data collection with existing simulation codes becomes impractical with the growing demand for large data sets.
In this paper, we introduce \textit{Cheetah}, a PyTorch-based high-speed differentiable linear-beam dynamics code. Cheetah is capable of accelerating beam dynamics simulations by multiple orders of magnitude through tensorised computation and several speed optimisation methods.
In the specific example of~\cite{kaiser2022learningbased}, this equates to a reduction in \ac{RL} training time from over \num{12} days when using the Ocelot simulation code~\cite{agapov2014ocelot} to just over \num{1} hour when using Cheetah.

At the same time, numerical optimisation is fast becoming an important tool for accelerator design, tuning, and model calibration~\cite{tomin2017ocelot,zhang2022badger}.
Advanced numerical optimisation methods like \ac{BO} have been used to achieve impressive results~\cite{roussel2023bayesian}. However, demands to solve optimisation problems of increasing dimensionality are growing, and \ac{BO} may struggle to efficiently optimise objective functions with more than a few dozen degrees of freedom~\cite{roussel2023bayesian}.
In the field of \ac{ML}, gradient-based optimisation has successfully been used to optimise up to \num{70} billion parameters~\cite{touvron2023llama,geminiteam2023gemini}.
However, computing gradients of complex models, like beam dynamics, using numerical or analytical methods are computationally expensive. Instead, automatic differentiation has found widespread adoption in machine learning for the fast computation of gradients.
\Ac{ML} frameworks, such as \textit{PyTorch}~\cite{paszke2019pytorch} and \textit{JAX}~\cite{jax2018github}, allow convenient and computationally cheap automatic differentiation to calculate the partial derivatives up to arbitrary orders for all the parameters using the chain rule.
Because Cheetah is constructed upon PyTorch, it provides built-in support for automatic differentiation to efficiently compute the gradients of the beam dynamics models it implements.
Hence, Cheetah makes optimisation over the large parameter spaces of accelerator facilities tractable beyond the number of parameters that can feasibly be optimised with the current state-of-the-art numerical optimisers.

Surrogate modelling of start-to-end accelerator systems utilising \acp{NN} is another active area of research~\cite{kaiser2023machine,edelen2017using,fujita2021physicsinformed}. Such surrogate models can be used to acquire offline models of processes in accelerator facilities or as a fast and differentiable stand-in for computationally expensive simulations.
Nevertheless, \ac{NN} are usually trained on start-to-end data, taking actuators as inputs and sensor values as output. This makes it difficult to reuse trained models for applications beyond those intended at the time of training.
Moreover, \ac{NN} surrogate models are not commonly designed to interface with beam dynamics simulators.
Conveniently, Cheetah is implemented using PyTorch, which is first and foremost an \ac{ML} framework. As a result, models implemented in Cheetah can be readily integrated with \ac{NN} surrogate models. This also means that gradient propagation from \ac{NN} surrogate models through Cheetah and vice versa works out-of-the-box.
With Cheetah, it is therefore possible to combine modular \ac{NN} surrogate models with physical beam dynamics simulations.
In particular, Cheetah provides a practical platform for integrating modular \ac{NN} surrogate models with handcrafted beam dynamics models, making the expensive-to-train surrogate models more reusable.

\begin{figure*}
    \centering
    \includegraphics[width=0.6\textwidth]{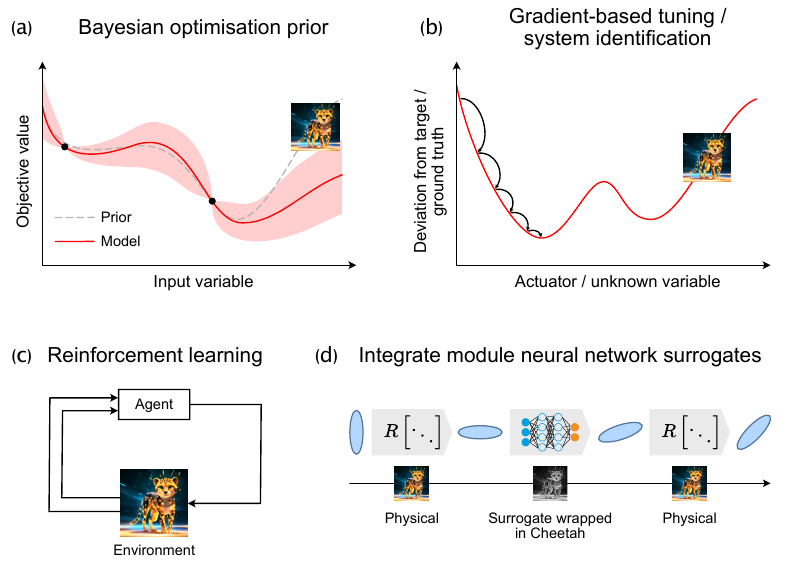}
    \caption{Overview of where Cheetah fits into the proposed applications, with the Cheetah logo marking its use as a component of these applications. (a) Cheetah is used as a physical prior for \ac{BO}. (b) Cheetah provides a differentiable beam dynamics model which can be used for accelerator tuning and system identification. (c) Cheetah enables the implementation of fast beam dynamics environments for training \ac{RL} agents. (d) Cheetah provides the infrastructure to seamlessly integrate modular \ac{NN} surrogate models with physical beam dynamics simulations.}
    \label{fig:cheetah_application_overview}
\end{figure*}

In the following, we introduce Cheetah and its inner workings in \cref{sec:implementing_linear_beam_dynamics_in_pytorch}, benchmarking its speed in \cref{sec:speed}. In the second half of this paper, we present five different application examples (also shown in context in \cref{fig:cheetah_application_overview}), taking advantage of Cheetah's speed for reinforcement learning in \cref{sec:highspeed_simulations_for_reinforcement_learning}, and using its differentiability for beam tuning in \cref{sec:gradientbased_beam_tuning} and system identification in \cref{sec:gradientbased_system_identification}, followed by an example using Cheetah as a \ac{BO} prior in \cref{sec:cheetah_prior_for_bayesian_optimisation}, and demonstrating how Cheetah may host modular \ac{NN} surrogate models in \cref{sec:integrating_modular_neural_network_surrogates_with_beam_dynamics_simulations}.

\subsection{Related work}\label{sec:related_work}

The field of programmatic beam dynamics modelling is very mature. There exist various well-established simulation codes for modelling beam dynamics in particle accelerators, e.g. ASTRA~\cite{floettmann1997astra}, Bmad~\cite{sagan2006bmad}, Elegant~\cite{borland2000elegant}, and MAD-X~\cite{cern1990madx}.
As Python has become increasingly popular in scientific computing, many of these have been augmented with Python adaptors. Further, the Ocelot~\cite{agapov2014ocelot}, Xsuite~\cite{iadarola2023xsuite}, and Bmad-X~\cite{gonzalez2023towards} simulation codes have been specifically developed directly in Python. Some calculations in Cheetah are based on those used in Ocelot.

\Acf{NN} surrogate modelling is also finding increased use to acquire fast and accurate models of complex beam behaviours~\cite{kaiser2023machine,edelen2017using}.
An \ac{NN} surrogate model is trained to infer the space charge field in a vacuum chamber cross-section using a physics-informed loss function including a partial differential equation with the Lorentz factor, elliptical bi-Gaussian charge density, and boundary condition in~\cite{fujita2021physicsinformed}.

An overview of opportunities for differentiable programming in particle physics instruments is given in~\cite{dorigo2023toward}.

Specialised handcrafted differentiable simulations have been constructed for various applications.
A handcrafted differentiable physics model is used as the discriminator in a \ac{GAN} setup to train an \ac{NN} to reconstruct time-domain measurements of X-ray pulses without labelled data in~\cite{ratner2021recovering}.
In~\cite{roussel2022differentiable,roussel2022applications} the hysteron density function of a Preisach model is fitted to accurately model hysteresis from experimental data.
In~\cite{roussel2022applications,roussel2023phase}, a differentiable beam dynamics simulation of a tomographic beamline is used to reconstruct phase space distributions from experimental screen images.
Simultaneous calibration of all detector parameters of a liquid argon time projection chamber using a differentiable simulation of the latter is performed in~\cite{gasiorowski2023differentiable}.
In~\cite{qiang2023differentiable}, a differentiable self-consistent space charge simulation model based on the truncated power series algebra (TPSA) is developed to speed up the simulated optimisation of accelerator design parameters under consideration of space charge induced effects.

A similar effort to Cheetah is pursued in~\cite{gonzalez2023towards}, where the authors introduce \textit{Bmad-X}, a library-agnostic differentiable particle tracking code written in Python based on Bmad. They demonstrate the application of Bmad-X on examples of beamline optimisation, model calibration, and phase-space reconstruction.
Bmad-X and Cheetah present very similar advantages, with both offering fast differentiable beam dynamics simulations.
However, they differ in some aspects.
In contrast to Cheetah, Bmad-X can be used with backend libraries other than PyTorch, while Cheetah has a stronger focus on fast computations and currently supports a larger number of lattice elements and conversions from other simulation codes.
Specifically, the goal of Cheetah is to bridge the gap between fast hand-crafted and data-driven particle accelerator models, streamlining their applications to various applications. As such, it aims to enable researchers to collect low-fidelity data fast and to use differentiable models to train \ac{ML} models or perform complex system identification.

An early preliminary version of Cheetah was first presented in~\cite{stein2022accelerating}. It was not yet designed to support automatic differentiation and lacked the majority of the features Cheetah now has.

\section{Fast differentiable linear beam dynamics in PyTorch}\label{sec:implementing_linear_beam_dynamics_in_pytorch}

The overarching goal in implementing Cheetah was to provide a differentiable beam dynamics code with improved speed
over existing simulation codes to be used for \ac{ML} applications.
Here, the conscious decision is made to trade accuracy to achieve these speed improvements. This means that Cheetah, while faster than existing codes, is lower fidelity than they are. With \ac{ML} applications this is a worthwhile trade-off. Methods like domain randomisation~\cite{tobin2017domain} enable \ac{NN} models trained on inaccurate simulated data to effectively generalise to the real-world domain. Moreover, initial training of an \ac{ML} model on cheap low-fidelity data followed by fine-tuning on high-fidelity data is a widely used method to speed up the training of \ac{ML} models.
At the same time, Cheetah is designed to integrate seamlessly with popular \ac{ML} tools. We intend for Cheetah to be used both as a tool in \ac{ML} applications, e.g. to support the training of neural network models, and as an application of \ac{ML} itself, e.g. through the integration of neural network models in its simulation pipeline.
Last but not least, our goal is to make Cheetah easy to use, easily extensible and follow best practices in its implementation with high-quality code.

To this end, Cheetah is implemented in the \textit{Python} programming language, which hosts an extensive \ac{ML} ecosystem and is widely used in scientific computing.
Cheetah employs the \textit{PyTorch}~\cite{paszke2019pytorch} framework. While the primary purpose of PyTorch is the implementation of \ac{ML} algorithms, its fast tensor compute capabilities, strong \ac{GPU} support and automatic differentiation features make it an ideal fit for fast parallel scientific computation.

To validate that Cheetah's models the physics of beam dynamics accurately and to ensure high code quality, Cheetah makes use of various \ac{CI} pipelines. Numerous tests are implemented to verify not only that Cheetah runs without errors, but also that its outcomes are physically plausible and match those computed by Ocelot~\cite{agapov2014ocelot,tomin2017ocelot}.
Automatic code formatting and linting are also used to enforce a high standard of readability and maintainability for Cheetah's code, while ensuring that implementations follow the best programming practices of PEP8 and minimising the incidence of elusive future errors.
The official GitHub repository~\cite{kaiser2021github} is set up with clear contribution guidelines and well maintained in an effort to foster future collaboration in the development of Cheetah.
To lower the barrier of entry and ease installation, stable versions of Cheetah are regularly deployed to \textit{PyPI}. Reference documentation and some use case examples for Cheetah are made available via \textit{Read the Docs}\footnote{\url{https://cheetah-accelerator.readthedocs.io}}.

\subsection{Beam tracking in Cheetah}

At its core, Cheetah is made up of two main object classes, \texttt{Beam} and \texttt{Element}, which provide implementations of charged particle beams and accelerator elements, such as magnets and drift sections, respectively. Both of these inherit from PyTorch's \texttt{Module}, allowing their parameters to be optimised with the tools provided by PyTorch when set to a \texttt{Parameter} instead of a \texttt{Tensor}.

Cheetah provides two ways to represent the beam, a \texttt{ParticleBeam} with coordinates of each macroparticle and a \texttt{ParameterBeam} with only statistical values representing the beam, both being a subclass of the \texttt{Beam}.
In \texttt{ParticleBeam}, each particle is represented by a seven-dimensional vector

\begin{equation}\label{eq:particle_representation}
    \bm{p} = \left( x, x', y, y', \tau, \delta, 1 \right),
\end{equation}

where $\{x,y\}$ are the horizontal and vertical positions, $\{x', y'\}$ are the slopes in trace space, $\tau$ the longitudinal displacement, and $\delta$ the momentum offset with respect to the nominal energy. The six-dimensional vector is expanded at the end, analogous to an affine space, allowing a coherent representation of transfer maps also for effects like magnet misalignments and thin-lens magnets.

For applications that require faster computations and do not require modelling of the bunch substructures, a second representation, the \texttt{ParameterBeam}, is used. It assumes a Gaussian beam and represents the entire beam by a seven-dimensional vector $\mu$ of the mean position in each dimension of the phase space and a covariance matrix $\Sigma$.

Furthermore, the \texttt{Beam} subclasses implemented in Cheetah offer convenient computation of various of their properties.
Both beam representations support generating Gaussian beams or being loaded from files saved by other particle tracking codes, a feature which is further discussed in~\cref{sec:integration_with_other_code}. In addition, \texttt{ParticleBeam} instances can be generated with regularly spaced macroparticles.

The \texttt{Element} class represents accelerator beamline elements, such as magnets, drift sections, or diagnostic instruments. Each subclass must implement a \texttt{track} method that transforms an incoming beam to an outgoing beam that was affected by the element. To add a new element to Cheetah, one simply inherits from \texttt{Element} and implements the \texttt{track} method. The \texttt{track} method can implement arbitrary computations from simple matrix multiplications for first-order tracking to more complex behaviours like higher-order transfer maps for non-linear elements, beam image computations for diagnostics screens, or neural network inference.
By default, Cheetah elements compute linear beam dynamics using an implementation of the linear transfer map $\transfermap_\text{Cheetah} \in \mathbb{R}^{7 \times 7}$ that is already provided

\begin{equation}\label{eq:linear_transfer_map}
    \transfermap_\text{Cheetah} = \left(
    \begin{array}{c|c}
            \raisebox{3pt}{\large\mbox{{$\transfermap_0$}}} & \vdots \\ \hline
            0 \cdots 0                                      & 1      \\
        \end{array}
    \right),
\end{equation}

with $\transfermap_0 \in \mathbb{R}^{6 \times 6}$ being the standard transfer matrix based on~\cite{brown1968first}.
For some elements more complex behaviours are already implemented in Cheetah. For example, the transverse motion in accelerating cavities is modelled according to~\cite{rosenzweig1994transverse}.
For the remainder of this paper, all transfer matrices $\transfermap$ are assumed to be of the form $\transfermap_\text{Cheetah}$.

To track a beam through an element with a transfer
matrix $R$, the default implementation either computes

\begin{equation}\label{eq:track_particle_beam}
    P_{\text{out}} = P_{\text{in}} R^\intercal
\end{equation}
for a \texttt{ParticleBeam} $P_{\text{in}} \in \mathbb{R}^{n \times 7}$ with $n$ macroparticles, or

\begin{equation}\label{eq:track_parameter_beam}
    \begin{aligned}
        \mu_{\text{out}}    & = R \mu_{\text{in}}                \\
        \Sigma_{\text{out}} & = R \Sigma_{\text{in}} R^\intercal
    \end{aligned}
\end{equation}
for a \texttt{ParameterBeam} with the characteristic parameters $\{\mu_\text{in}, \Sigma_\text{in} \}$.

For elements that only implement linear beam dynamics, it is therefore sufficient to implement a \texttt{transfer\_map} method returning a first order transfer matrix $R$ for the element.
At the time of writing this paper, Cheetah has support for drift sections, dipole magnets with adjustable face angles (e.g. SBends and RBends), thin-lens corrector magnets, quadrupole magnets, cavities, \acp{BPM}, markers, diagnostic screen stations, apertures, solenoid magnets, and elements with custom transfer maps.
In addition, Cheetah provides a special \texttt{Segment} subclass of \texttt{Element}. It represents a sequential lattice of accelerator components and supports nesting of other smaller \texttt{Segment} elements.

We continue to extend Cheetah with new elements and features. In a further community-driven effort, users of Cheetah can add new features, such as elements, physical processes, and specialised transfer maps, according to task specific requirements.

\subsection{Speed optimisation}

Cheetah achieves its speed through several automatic and opt-in optimisations. First of these is the use of PyTorch, which is itself implemented in \textit{C++} and \ac{CUDA} and is well-optimised thanks to widespread community support.
PyTorch holds a key speed advantage over established packages like NumPy in its built-in ability to run on \acp{GPU} supporting \ac{CUDA} or \ac{MPS}, which can provide significant speed improvements for massively parallel computations such as single particle tracking.

\begin{figure}
    \centering
    \includegraphics{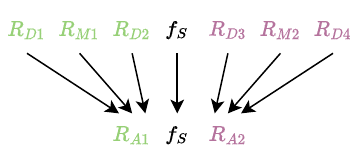}
    \caption{Visualisation of a simple example for transfer map reduction. The tracking function of the screen is denoted by $f_S$. It cannot be reduced along with other transfer maps. The transfer maps drift sections and magnets upstream of the screen $\left\{ \transfermap_{D1}, \transfermap_{M1}, \transfermap_{D2} \right\}$ and downstream of the screen $\left\{ \transfermap_{D3}, \transfermap_{M2}, \transfermap_{D4} \right\}$ can be reduced to two transfer maps $\transfermap_{A1}$ and $\transfermap_{A2}$, one on each end of the screen.}
    \label{fig:transfer_map_reduction}
\end{figure}

Moreover, Cheetah automatically identifies sequences of elements that can have their transfer matrices combined. We refer to this optimisation as \textit{dynamic transfer map reduction}. For example, if a \texttt{Segment} is made up of an alternating sequence of dipole magnets and drift sections following linear beam dynamics, followed by an active diagnostic screen station and another sequence of alternating dipole magnets and drift sections, Cheetah will automatically recognise that the transfer matrices $\left\{ \transfermap_{M1}, \transfermap_{D1}, \transfermap_{M2}, \transfermap_{D2} \dots \right\}$ of the elements upstream of the screen station can be combined into a single transfer matrix $\transfermap_{\text{upstream screen}}$, and that the same can be done for the sequence of elements downstream of the screen station.
A simple example following this description is shown in \cref{fig:transfer_map_reduction}.
This optimisation can be influenced by the user to some extent. Some elements, such as diagnostic screen stations and \acp{BPM}, support being activated or deactivated, based on whether the user intends to use their functions. Cheetah makes transfer map reduction decisions based on the activation status of these elements. Other elements such as cavities, automatically determine whether they can be optimised through transfer map reduction. Cavities, for example, produce a drift transfer matrix when inactive, which is automatically combined with other transfer maps, but do not take part in transfer map reduction when they are active and have more complex effects on the beam.
Because transfer map reduction does not always need to be performed every time a beam is tracked through a \texttt{Segment},
Cheetah provides an opt-in variant of the same optimisation, where the user can tell Cheetah which elements may be changed in the future. All other elements are then frozen, allowing Cheetah to perform \textit{static transfer map reduction}. This optimisation can be very effective when only a few parameters are changed between consecutive simulations on large lattices.

In a similar vein, Cheetah can be commanded to find all inactive elements that are effectively drift sections, and replace them with actual drift sections, which are generally faster to compute. In addition, drift sections in Cheetah are pure linear beam dynamics elements, meaning they can be merged with adjacent linear beam dynamics elements, either dynamically and on-the-fly, or statically before tracking.

Especially when combining these optimisations, they can significantly speed up computations in Cheetah. Critically, all the implemented optimisations maintain the differentiability and correct gradients of the models.

\subsection{Integration with other codes}\label{sec:integration_with_other_code}

To facilitate the quick adoption of Cheetah, the ability to load beams and lattices, especially from other particle tracking codes, is crucial. Cheetah's default lattice exchange format is an adapted variant of the interoperable lattice exchange format \textit{LatticeJSON}~\cite{andreas2019latticejson}. Based on the standard JSON format, this makes reading and writing of lattice files which are compatible with Cheetah straightforward in any programming language.
Cheetah's modular and simple architecture further simplifies the implementation of converters from other lattice and beam exchange formats.
Cheetah currently supports loading beams from ASTRA, lattices from Bmad, and both beams and lattices from Ocelot.

\subsection{Speed benchmarks}\label{sec:speed}

In this section, we benchmark the execution speed of Cheetah with other simulation codes on the same tracking tasks. The lattice considered for the benchmark is the \textit{Experimental Area} section of the \textit{ARES} accelerator~\cite{panofski2021commissioning,burkart2022the} at DESY, which is further investigated in some of the use case examples in~\cref{sec:use_case_example}. The section is in total \qty{2.05}{\meter} long, consisting of three quadrupole magnets, two corrector magnets, and drift sections in between.
The benchmarks were run on two different computing platforms to account for the potential advantages of different hardware. Firstly, we ran simulations on a laptop with an Apple M1 Pro with \num{10} CPU Cores and \qty{32}{\giga\byte} of RAM. Secondly, we considered a \ac{HPC} cluster node with two AMD EPYC 7643 having a combined \num{192} cores, \qty{1024}{\giga\byte} of RAM, and \num{4} Nvidia A100 \acp{GPU}, each having \qty{80}{\giga\byte} of VRAM. Both the \ac{CPU} and the \ac{GPU} were considered with Cheetah on the cluster node. Note that at the time of writing, Cheetah can only use one GPU at any time.
Simulation times were averaged over multiple runs using Python's \textit{timeit} package.
Cheetah was run in multiple different configurations: tracking a \texttt{ParameterBeam}, tracking a \texttt{ParticleBeam} on CPU, and tracking a \texttt{ParticleBeam} on GPU. For all the configurations, we benchmarked with and without the opt-in lattice optimisations.
We further compared Ocelot and ASTRA~\cite{floettmann1997astra} with and without space charge. Parallel ASTRA was run using \num{8} performance cores on M1 Pro and \num{48} cores on EPYC 7643, which we found to be the fastest configurations for this particular benchmark.
In addition, we consider Bmad-X with a NumPy backend and Xsuite for the benchmarks.
The results of the speed benchmark are listed in \cref{tab:benchmark_speed}. For Cheetah with \texttt{ParticleBeam} and other simulation codes, a beam with \num{100000} macroparticles is used.
The benchmarks were run with a pre-release version of Cheetah v0.6.2.

We find that Cheetah can compute the benchmarked simulation setup up to \num{8} orders of magnitude faster than the other benchmarked simulation codes.
In particular, Cheetah is about \num{5500} times faster than the fastest ASTRA setup without space charge on the ARM laptop.
The fastest setup of Ocelot is outperformed by Cheetah by over \num{9000} times on the same device.
In our benchmarks, Cheetah also achieves computational speeds around \num{1900} times faster than the already very fast Bmad-X.
Xsuite achieves speeds comparable to Cheetah without Cheetah's opt-in optimisations turned on, but Cheetah is up to two orders of magnitude faster when opt-in optimisations are used.
Remember that these speed advantages of Cheetah by design come at the cost of accuracy, where higher-order effects, collective effects, and others are left out by default in order to achieve the reported speeds.
We further find that in our benchmark Cheetah's \texttt{ParameterBeam} is tracked between \num{2} and \num{40} times faster than the same \texttt{ParticleBeam}.
GPU acceleration is a sensible choice only with \texttt{ParticleBeam}, though it is not guaranteed to improve compute times. While we did observe \num{8} times faster simulation with lattice optimisations turned on, simulations slowed down by a factor of almost \num{1.6} without them. 
This is the result of the benchmark beam tracking \num{100000} particles. In this case, the overhead induced by sending instructions and data to the GPU outweighs the performance benefits of highly-parallel computation. On the other hand, when the number of tracked particles is increased to \num{10000000}, tracking with optimisations turned on takes \qty{37.5}{\milli\second} on \ac{CPU} and \qty{998}{\micro\second} on \ac{GPU}. With optimisations turned off, Cheetah tracks the same beam in \qty{37.5}{\milli\second} on \ac{CPU} and \qty{5.36}{\milli\second} on \ac{GPU}. This is a significant improvement, demonstrating the advantages of \ac{GPU} acceleration in Cheetah.
Moreover, we find that in our benchmarks, opt-in optimisations yielded up to \num{38} times faster execution on \ac{CPU} and up to \num{51} times faster execution on \ac{GPU}. Note that the opt-in optimisations benchmarked here are the most extreme case, taking maximum advantage of the optimisations to demonstrate Cheetah at its fastest and at its slowest. In real-world use of opt-in optimisations, we expect results to be slightly worse than the optimised cases showcased here, as some user-defined exceptions might reduce the effectiveness of Cheetah's optimisations.

\begin{table*}
    \centering
    \caption{Step computation times of simulation codes in milliseconds}
    \label{tab:benchmark_speed}
    \begin{tabular}{llrr}
        \toprule
        \textbf{Code}  & \textbf{Comment}                           & \textbf{Laptop} & \textbf{\ac{HPC} node} \\
        \midrule
        ASTRA          & space charge                               & \num{264000.00} & \num{3605000.00}       \\
                       & no space charge                            & \num{109000.00} & \num{183000.00}        \\
        Parallel ASTRA & space charge                               & \num{39000.00}  & \num{17300.00}         \\
                       & no space charge                            & \num{16900.00}  & \num{12600.00}         \\
        Ocelot         & space charge                               & \num{22100.00}  & \num{21700.00}         \\
                       & no space charge                            & \num{182.00}    & \num{119.00}           \\
        Bmad-X         &                                            & \num{40.50}     & \num{74.30}            \\
        Xsuite         & CPU                                        & \num{0.81}      & \num{2.82}             \\
                       & GPU                                        & -               & \num{0.57}             \\
        Cheetah        & \texttt{ParticleBeam}                      & \num{1.60}      & \num{2.95}             \\
                       & \texttt{ParticleBeam} + optimisation       & \num{0.79}      & \num{0.72}             \\
                       & \texttt{ParticleBeam} + GPU                & -               & \num{4.63}             \\
                       & \texttt{ParticleBeam} + optimisation + GPU & -               & \num{0.09}             \\
                       & \texttt{ParameterBeam}                     & \num{0.76}      & \num{1.29}             \\
                       & \texttt{ParameterBeam} + optimisation      & \num{0.02}      & \num{0.04}             \\
        \bottomrule
    \end{tabular}
\end{table*}

\section{Use case examples}\label{sec:use_case_example}

In the following we would like to demonstrate in five examples how Cheetah might be used and what it is capable of.
In \cref{sec:highspeed_simulations_for_reinforcement_learning}, we demonstrate on the example of recently published work, how Cheetah can be used to enable fast reinforcement learning in simulation for policies that transfer well to the real world. This is followed by examples of using Cheetah's automatic differentiation features to perform gradient-based accelerator tuning in \cref{sec:gradientbased_beam_tuning} and gradient-based system identification in \cref{sec:gradientbased_system_identification}. In \cref{sec:cheetah_prior_for_bayesian_optimisation}, we show Cheetah's utility as a physics-based prior in the context of Bayesian optimisation. At last, we demonstrate Cheetah's suitability for an extension by modular element neural network surrogate models in \cref{sec:integrating_modular_neural_network_surrogates_with_beam_dynamics_simulations}.
The following is not an exhaustive list of applications for Cheetah. We believe that as Cheetah is adopted, users will find many more problems it can solve.

\subsection{High-speed simulations for reinforcement learning}\label{sec:highspeed_simulations_for_reinforcement_learning}

In recent work, Cheetah played a key role in the successful training of a neural network policy for tuning the transverse beam properties in a particle accelerator through the method of \ac{RL}~\cite{eichler2021first,kaiser2022learningbased,kaiser2023learning} -- so-called \acf{RLO}.
Specifically, this work considers a tuning task in the \acf{EA} beamline section at the \textit{ARES} accelerator.
The \ac{EA} is made up of a sequence $\left\{ \quadrupole_1, \quadrupole_2, \steerer_v, \quadrupole_3, \steerer_h \right\}$ of two quadrupole magnets, followed by a vertical dipole magnet, a third quadrupole magnet and a horizontal dipole magnet. These magnets allow for the tuning of the transverse beam properties $\left( \mu_x, \sigma_x, \mu_y, \sigma_y \right)$, i.e.~position and size in the horizontal and vertical dimensions.
These properties are measured with a diagnostic screen station downstream of the magnets.
At ARES, transverse beam parameter tuning is commonly performed in preparations for experiments in an experimental vacuum chamber installed downstream of the \ac{EA}.
The goal of the transverse beam tuning task in the \ac{EA} is to find the magnet settings \mbox{$\actuators = \left(\quadstrength_{\quadrupole_1}, \quadstrength_{\quadrupole_2}, \kick_{\steerer_v}, \quadstrength_{\quadrupole_3}, \kick_{\steerer_h} \right)$} that minimise the difference between the beam parameters observed on the screen $\measuredbeam = \left( \mu_x, \sigma_x, \mu_y, \sigma_y \right)$ and some target beam parameters $\targetbeam = \left( \mu_x', \sigma_x', \mu_y', \sigma_y' \right)$ set by a human operator.
In the \ac{EA}, the beam entering that section $\incomingbeam$ and the transverse misalignments $\misalignments$ of components like quadrupoles and the screen are not known, making this the transverse beam tuning task more challenging to solve.
To date, transverse beam tuning is mostly solved manually by experienced human operators, which requires a lot of time and makes it difficult to reproduce results.

\begin{figure}
    \centering
    \includegraphics[width=\columnwidth]{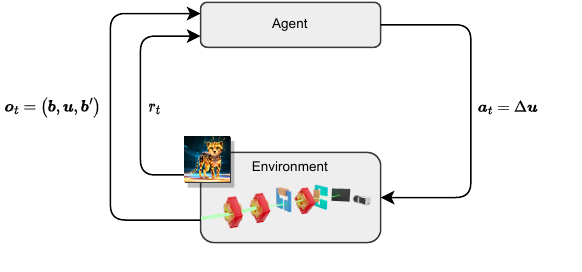}
    \caption{Flowchart of the \ac{RL} loop for the ARES \ac{EA} transverse tuning task. The environment -- during training defined in terms of Cheetah -- outputs an observation $\observation_t$ and a $\reward_t$ based on the previous action $\action_{t-1}$.
        The agent then computes a new action $\action_t$ using the neural network policy.
        The new action is applied to the environment and results in the next observation $\observation_{t+1}$ and reward $\reward_{t+1}$.}
    \label{fig:ares_rl_loop}
\end{figure}

In order to solve this beam tuning task utilising \ac{RLO},
a task-specific \ac{RL} loop is defined as shown in \cref{fig:ares_rl_loop}.
In this loop, the accelerator environment is implemented using Cheetah at the time of training and then replaced with the real ARES accelerator at the time of application. The Python package \textit{Gymnasium}~\cite{towers2023gymnasium} (the successor to the previously popular \textit{OpenAI Gym}~\cite{brockman2016openai}) is used to define the environment.
A \ac{MLP} of two hidden layers is used as a policy model. Each layer has a width of \num{64} neurons and uses a \ac{ReLU} activation. The policy takes as input the normalised observed beam $\measuredbeam$, the currently set quadrupole strengths and deflection angles of the magnets $\actuators$, and the target beam parameters $\targetbeam$.
Its output is defined as normalised changes to the magnet settings $\action_t = \Delta \bm{u}$.
The rewards and observations are normalised using a running average during the training. The actions are normalised to the action spaces, which is [-3, 3]~\unit{\per\meter\squared} for quadrupole strengths, [-0.6, 0.6] \unit{\milli\radian} for vertical steering magnet, and [-0.3, 0.3] \unit{\milli\radian} for horizontal steering magnet. The different action ranges of vertical and horizontal magnets are chosen so that they have approximately the same steering effect at the position of the diagnostics screen.

To train the policy, the \ac{TD3}~\cite{fujimoto2018addressing} algorithm is used for its relative training sample efficiency among model-free RL algorithms. Specifically, we employ the implementation provided by the \textit{Stable Baselines3}~\cite{raffin2021stable} Python package.
As originally introduced to the field of \ac{RL} for accelerators in~\cite{kaiser2022learningbased}, domain randomisation~\cite{tobin2017domain} is performed during training by sampling magnet and screen misalignments, as well as the incoming beam parameters and the target beam from a uniform distribution at the start of each rollout episode.
We define the reward for transverse beam parameter tuning as

\begin{equation}
    R\left(\state_t, \action_t\right) = \begin{cases}
        \hat{R} \left( \state_t, \action_t \right)       & \text{if $\hat{R} \left(\state_t, \action_t\right) > 0$} \\
        2 \cdot \hat{R} \left(\state_t, \action_t\right) & \text{otherwise}\,
    \end{cases}
\end{equation}

with $\hat{R} \left(\state_t, \action_t\right) = O\left(\actuators_t\right) - O\left(\actuators_{t+1}\right)$, where the objective function $O\left(\actuators_t\right)$ is the logarithmic and weighted difference between the observed and target beams

\begin{equation}
    O\left(\actuators_t\right) \coloneqq \ln \frac{1}{4} \sum_{i=1}^{4} \weights^{(i)} \left| \measuredbeam^{(i)} - \targetbeam^{(i)} \right| \text{.}
\end{equation}

A weight vector $\weights = \left( \num{1}, \num{2}, \num{1}, \num{2} \right)$ was chosen for the final training.

The policy is trained over a total of \num{6000000} interactions with the Cheetah environment. One interaction with the real environment involves sending new set points to the magnet power supplies, waiting for the power supplies to finish ramping to their set points, and taking multiple images of the diagnostic screen with the beam turned on and with the beam turned off.
Altogether this process takes ca. \qtyrange{10}{20}{\second}.
Consequently, training on the real ARES accelerator would take about \num{3} years of continuous beam time.
With beam time a scarce resource, such a long training is infeasible.
Fast simulations like Cheetah can be computed faster than real time and enable us to collect the equivalent of many years of real time experience in much more feasible time frames. As was shown in \cref{sec:speed}, Cheetah is especially fast, allowing for the equivalent of \num{3} years of experience to be collected in just \qty{27}{\minute}.
Other \ac{RL} algorithms such as \ac{PPO} also allow for parallel rollouts on multiple environments. Using a simulation like Cheetah means that experience could be collected even faster in this case, while we usually do not have access to multiple of the same accelerator for training.

\begin{figure}
    \centering
    \includegraphics{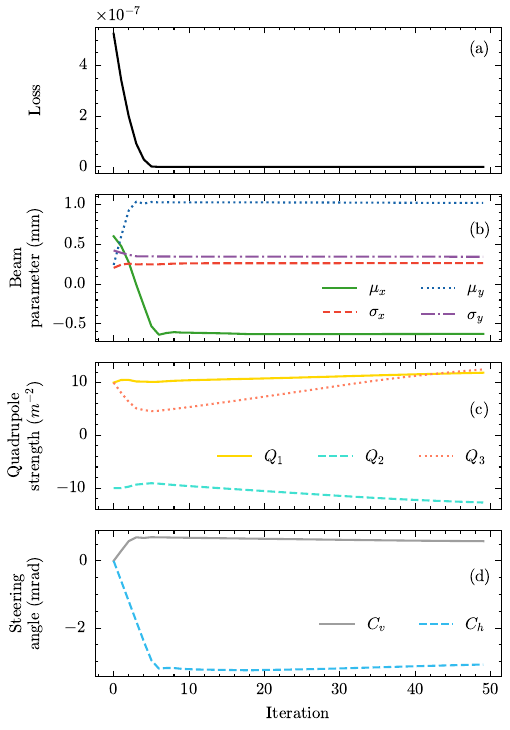}
    \caption{An \ac{NN} policy trained with \ac{RL} tuning of the transverse beam parameters in the ARES \ac{EA}. (a) The \ac{MSE} loss development over parameter update iterations. (b) Beam parameters on the diagnostics screen over parameter update iterations. (c) and (d) quadrupole and dipole magnet settings over parameter update iterations.}
    \label{fig:reinforcement_learning_example_history}
\end{figure}

Despite having been trained using a comparatively simple simulation and deployed to the real world zero-shot without fine-tuning, our \ac{RL} policy successfully tunes the transverse beam parameters on the real accelerator, finding magnet settings that achieve beam parameters closer to the target than those found by other state-of-the-art black-box optimisation algorithms. Moreover, the trained policy converges on these magnet settings in just a few samples, tuning the beam in less wall time than human operators while achieving comparable results.
An example of a trained policy from~\cite{kaiser2023learning} tuning the transverse beam parameters in the \ac{EA} is shown in \cref{fig:reinforcement_learning_example_history}.
Here it can be observed that the target transverse beam parameters of $\targetbeam = \left( \qty{-0.61}{\milli\meter}, \qty{0.26}{\milli\meter}, \qty{1.03}{\milli\meter}, \qty{0.35}{\milli\meter} \right)$ are reached after about \num{6} steps.
For more detailed results and discussions, please refer to~\cite{kaiser2022learningbased,kaiser2023learning}.

\subsection{Gradient-based beam tuning and lattice optimisation}\label{sec:gradientbased_beam_tuning}

In the field of particle accelerators, there are various optimisation tasks, ranging from lattice optimisation in the design phase of a facility to tuning actuators at run time.
In some cases, these tasks have too many degrees of freedom to be feasibly solvable by black-box optimisation algorithms, such as \ac{BO} or \ac{RLO}. However, their underlying function and its parameters may be known. In such cases, gradient-based optimisation may be used. The latter has well-understood convergence properties and extensive tooling for it has been developed in the field of \ac{ML}.
Using gradient-based tuning on a model of an accelerator can help find good setups without the need for beam time. Even in cases where there exists a model mismatch, this offline optimisation approach can provide good starting points close to the optimum, which can then be further optimised online.
Further, Cheetah can be used to reduce model mismatches through gradient-based system identification as is described in \cref{sec:gradientbased_system_identification}.

In this example, we consider the same transverse beam tuning task as in \cref{sec:highspeed_simulations_for_reinforcement_learning}. In contrast to before, we assume that unobserved properties, such as the incoming beam and the beamline components' misalignments, are known. This may be the case in simulations during the design stage of an accelerator, if an accelerator is known to deviate very little from its design parameters, or if system identification like in \cref{sec:gradientbased_system_identification} was performed ahead of time.

The ARES lattice is loaded as a Cheetah \texttt{Segment}. Because Cheetah defines segments as PyTorch \texttt{Module}, all that is needed for PyTorch to automatically compute the gradient of the ARES \ac{EA} with respect to the five magnet settings, is to define the latter as PyTorch \texttt{Parameter}. A fixed incoming beam is tracked through the \ac{EA} Cheetah \texttt{Segment}. The resulting beam parameters can then be read from the diagnostic screen station at its end, and a \ac{MSE} loss, defined as

\begin{equation}
    \frac{1}{4} \sum_{i=1}^{4} \left( \measuredbeam^{(i)} - \targetbeam^{(i)} \right) ^ 2 \text{,}
\end{equation}

with $\targetbeam$ being the target beam and $\measuredbeam$ the currently observed beam at the screen station, can be computed, where $\measuredbeam^{(i)}$ denotes the $i$-th element of $\measuredbeam$. PyTorch's automatic differentiation features can then be used to compute the gradient of the particle tracking and transverse beam parameter loss function with respect to the magnet settings

\begin{equation}
    \measuredbeam = f_{\text{EA}} \left( \actuators \mid \misalignments, \incomingbeam  \right) \text{.}
\end{equation}

Adam~\cite{kingma2014adam}, a variant of \ac{SGD} is then used to compute the updates to the magnet settings based on the computed gradient.

\begin{figure}
    \centering
    \includegraphics{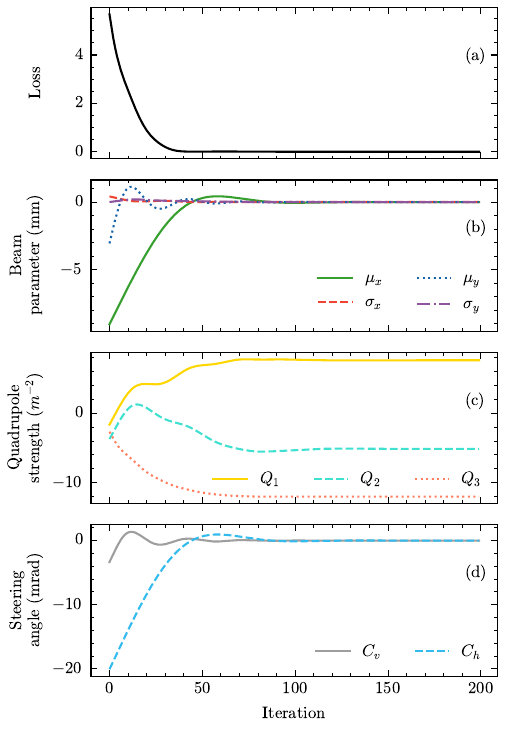}
    \caption{Gradient-based tuning example of the transverse beam parameters in the ARES \ac{EA}. (a) The loss development over parameter update iterations. (b) Beam parameters on the diagnostics screen over parameter update iterations. (c) and (d) quadrupole and dipole magnet settings over parameter update iterations.}
    \label{fig:tuning_convergence_example}
\end{figure}

However, as is, this simple setup would result in unstable convergence. This is because the magnet settings are on very different scales, with the maximum quadrupole setting at \qty{72}{\per\meter\squared} and a maximum dipole magnet setting of \qty{6.2}{\milli\radian}. To address this, the magnet settings are normalised, i.e. scaled to normally fall into the range $\left[-1, 1 \right]$. With Cheetah, this is easily achieved by wrapping the ARES \ac{EA} \texttt{Segment} in an outer PyTorch \texttt{Module} with the normalised magnet settings as the PyTorch \texttt{Parameter}.
On every call to the module's \texttt{forward} method, the segment's magnet settings $\actuators$ are set to

\begin{equation}
    \actuators = \normalizedactuators \odot \normalizationscalevector \text{,}
\end{equation}

i.e.~the element-wise product of the normalised actuator parameters $\normalizedactuators$ and the scaling factors for each respective actuator component $\normalizationscalevector$.
For the presented case study we use $\normalizationscalevector = \left( \qty{5}{\per\meter\squared}, \qty{5}{\per\meter\squared}, \qty{6.2}{\milli\radian}, \qty{5}{\per\meter\squared}, \qty{6.2}{\milli\radian} \right)$.
Note that for the quadrupole magnets, the scaling factors are chosen to be smaller than the physical limits of the real magnets so that they represent the commonly used operational ranges of these magnets at ARES more accurately.

The resulting convergence of the magnet settings can be seen in \cref{fig:tuning_convergence_example}.
In the shown example, the target beam parameters are $\targetbeam = \left( \qty{0.0}{\micro\meter}, \qty{0.0}{\micro\meter}, \qty{0.0}{\micro\meter}, \qty{0.0}{\micro\meter} \right)$.
We observe that the final magnet settings result in the desired centred and focused beam.
The absolute deviation of the observed transverse beam parameters to the target transverse beam parameters is $\left|\Delta\beampositionx\right| = \qty{0.33}{\micro\meter}$, $\left|\Delta\beamwidth\right| = \qty{6.66}{\micro\meter}$, $\left|\Delta\beampositiony\right| = \qty{0.07}{\micro\meter}$, and $\left|\Delta\beamheight\right| = \qty{0.85}{\micro\meter}$, resulting in a \ac{MAE} of \qty{1.98}{\micro\meter}.
Convergence is generally smooth, with all five magnets converging on their final settings after about \num{90} gradient steps.
Note that the hyperparameters for this example were not tuned and better results may be possible.

\subsection{Gradient-based system identification and virtual diagnostics}\label{sec:gradientbased_system_identification}

A common challenge with accelerators is that some properties of the beam or the accelerator hardware itself are not observable. Finding these properties usually requires multiple samples at different system states, ideally collected in a structured manner such as a grid scan for best results. Using these samples to reconstruct the hidden properties constitutes an inverse problem. Inverse problems are notoriously difficult to solve. Performing structured measurements to identify hidden properties of an accelerator necessitates an interruption of beam delivery, making it a costly measurement that is only performed if strictly needed.

Here we consider a system identification task in the ARES \ac{EA}. There are two unknowns in the \ac{EA}: The incoming beam and the misalignments of the quadrupoles. For this example, we aim to identify the misalignments of the quadrupoles. Knowing these can help tune the accelerator, for example by inserting the found misalignments into the model used in \cref{sec:gradientbased_beam_tuning}, or by using them to better align the quadrupoles and thereby reduce the dipole effect they have on the beam when used for focusing.

The gradients are computed with respect to the misalignments, i.e. the horizontal and vertical displacements of the magnets. Similar to the tuning example, the misalignments are normalised by wrapping the \ac{EA} \texttt{Segment} in a PyTorch Module that holds normalised misalignment $\misalignments_\text{normed}$ Parameters such that

\begin{equation}
    \measuredbeam = f_{\text{EA}} \left( \misalignments_\text{normed} | \incomingbeam, \actuators \right) \text{.}
\end{equation}

The resulting optimisation problem is defined as

\begin{equation}
    \text{min}_{\misalignments} O \left( \misalignments \right) = \left( \mu_x - \mu_x' \right) + \left( \mu_y - \mu_y' \right)
\end{equation}
to find the misalignments such that, when these misalignments are assumed in the model, the beam positions predicted on the screen best match the experimental measurements. For this example, we use parasitically acquired measurements from other unrelated experiments. This effectively enables zero-shot system identification. The specific data used here was collected during evaluations of the \ac{RLO} policies in \cite{kaiser2023learning}, which was also referenced in the example in \cref{sec:highspeed_simulations_for_reinforcement_learning}.

\begin{figure*}
    \centering
    \includegraphics{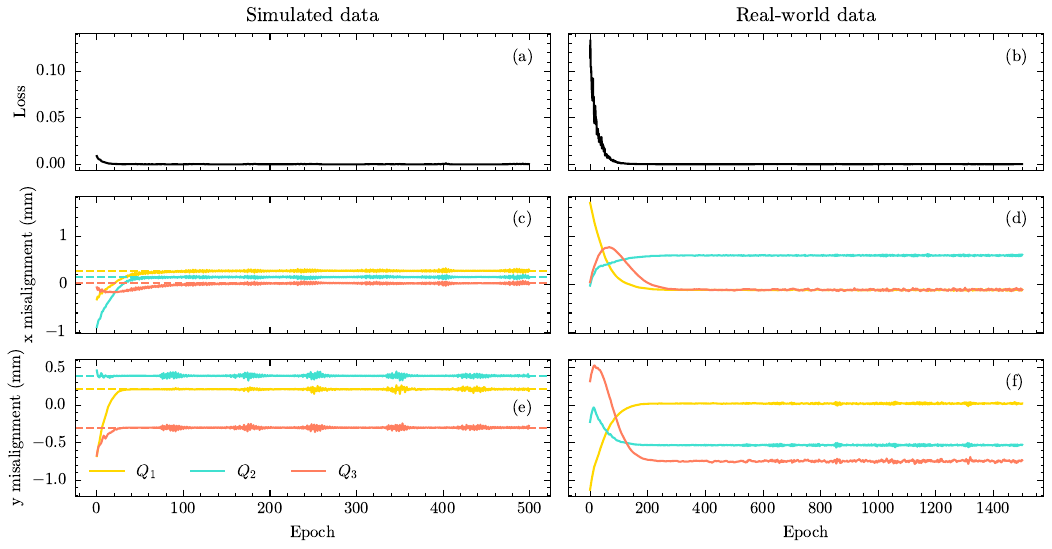}
    \caption{Two examples of gradient-based quadrupole misalignment identification. For the first example, using parasitically collected simulated data, we show in (a) the loss, (c) the horizontal misalignments, and (e) the vertical misalignments over epochs. Dashed lines signify the ground truth misalignments, which are known in simulation. We show in (b) the loss, (d) the horizontal misalignments, and (f) the vertical misalignments for the second example using real-world parasitically collected data.}
    \label{fig:system_id_combined}
\end{figure*}

We first start with data collected in simulations, where the misalignments are known. This allows us to verify that the gradient-based optimisation arrives at correct misalignments. Because we are considering simulated data, the correct incoming beam is also known and may be assumed in the Cheetah model. As can be seen in \cref{fig:system_id_combined} (a), (c), and (e), the reconstruction arrives at the correct misalignments.

Now trusting the misalignment reconstruction, the latter may be tried on data collected from a real-world experiment.
With real-world data, however, the incoming beam is often unknown. Therefore, a sensible assumption on the incoming beam must be made. In the ARES \ac{EA}, the goal is to reduce the misalignments of the quadrupole magnets with respect to the design orbit, which is the centre of the beam pipe in most cases.
We therefore assume that the incoming beam position and momentum are at zero, i.e. we consider the observed orbit to be the design orbit.
Doing so effectively sets the origin of the Cheetah model to the observed orbit in the data, resulting in misalignments being measured as deviations from that orbit. Other properties such as the beam size and energy only marginally affect this particular system identification setup and are therefore not considered.
As can be seen in \cref{fig:system_id_combined} (b), (d), and (f), the reconstruction appears to also perform well and smoothly arrive at sensible results under these conditions. However, the results cannot be checked against the ground truth this time, because the ground truth cannot be known.

\subsection{Physics-based prior mean for Bayesian optimisation}\label{sec:cheetah_prior_for_bayesian_optimisation}

Thanks to its speed, Cheetah can provide fast predictions of the beam parameters and guide optimisation algorithms during online tuning tasks, ultimately boosting their performance.
One particular use case is \ac{BO}, which utilises a \ac{GP} model to build a surrogate model of the observed data and efficiently optimise the objective function.
However, when dealing with high-dimensional tasks, \ac{BO} tends to over-explore the parameter space to find the global optimum, inevitably increasing the required number of samples~\cite{shahriari2016taking}. This limits the tasks which are solvable with classical BO algorithms to those that have less than a few dozen of input parameters~\cite{roussel2023bayesian}.
Recent studies have shown that the convergence speed of \ac{BO} can be significantly improved by incorporating prior knowledge of the accelerator system into the \ac{GP} model, for example by including the correlation of quadrupole magnets into the \ac{GP} covariance~\cite{duris2020bayesian} or using an \ac{NN} surrogate model as the prior mean function for the \ac{GP}~\cite{xu2022neural}.
In the case of an \ac{NN} surrogate, it should be accurate enough, as a wrong prior will only hamper the performance of \ac{BO}. However, training such an \ac{NN} model requires many samples either from simulation or real measurements, which are often not readily available.
Cheetah becomes a promising alternative for the prior mean function for BO due to its fast inference time. In addition, Cheetah's differentiability allows efficient acquisition function optimisation using gradient descent methods in modern \ac{BO} packages like \textit{BoTorch}~\cite{balandat2020botorch}.

In the following, we demonstrate the usage of Cheetah as a prior mean for BO in a beam-focusing task. The investigated lattice segment is a FODO cell consisting of two quadrupole magnets $\{\quadrupole_1, \quadrupole_2\}$ and a diagnostic screen at the end. The objective is to minimise the mean beam size measured at the screen

\begin{equation}
    O \coloneqq \frac{1}{2} (\left|\sigma_{x} \right| + \left|\sigma_{y} \right|)
\end{equation}
by changing the quadrupole strengths $\{ \quadstrength_{\quadrupole_1}, \quadstrength_{\quadrupole_2}\}$, where $\sigma_{x}$ and $\sigma_{y}$ denote the horizontal and vertical beam sizes respectively.

\begin{figure}
    \centering
    \includegraphics[width=\linewidth]{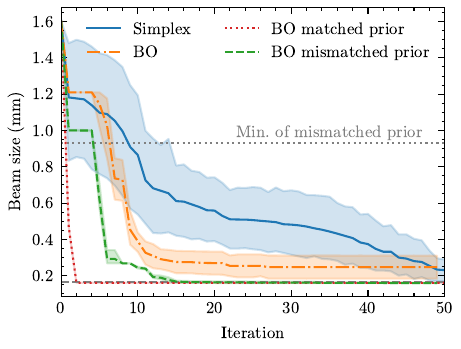}
    \caption{Optimisation results of a beam focusing task using Nelder-Mead simplex (blue), Bayesian optimisation with a constant mean function (orange), and BO with a Cheetah \texttt{Segment} as the prior mean matched to the task (red) or mismatched (green). The results are averaged over \num{10} runs for each algorithm and shaded regions represent one standard deviation. The dotted line shows the minimum objective of the prior mean in the mismatched lattice and the dashed grey line denotes the true minimum for each task obtained from grid scans.}
    \label{fig:bo_cheetah_prior}
\end{figure}

We evaluated the performance of \ac{BO} with a Cheetah simulation model as a prior mean function and used \ac{BO} with a constant prior mean and Nelder-Mead simplex as baselines.
All algorithms are implemented based on the \textit{Xopt} package~\cite{roussel2023xopt}. Both \ac{BO} variants use the Mat\'ern-5/2 kernel and \ac{UCB} acquisition function with $\beta=2.0$, which is a standard choice of hyperparameters for \ac{BO} applications. Each algorithm was repeated \num{10} times starting from the same detuned setting and the averaged results with one standard deviation are shown in \cref{fig:bo_cheetah_prior}.
In the case of the prior mean matched to the tuning task, \ac{BO} with prior could immediately find the global minimum without exploring much of the parameter space.
We then changed the lattice distances so that the \ac{BO} prior mean is mismatched to the ground truth of the tuning task. \Ac{BO} with prior first sampled around the minimum predicted by the prior mean, which is denoted by the dotted line in \cref{fig:bo_cheetah_prior}(b). Since there was a difference between the predicted and observed beam sizes, it continued exploring the parameter space and subsequently converged to the minimum.

In both cases, \ac{BO} with prior converged to the true minimum within \num{15} steps and was more sample-efficient than \ac{BO} with a constant mean prior and simplex algorithm even for the two-dimensional task.
This is expected to have a larger impact on higher-dimensional tasks as the parameter space grows exponentially.
Furthermore, when \ac{BO} with a Cheetah prior mean is applied to real-world tasks, one can use system identification to determine the mismatch between the simulated and the real accelerator using the obtained data parasitically, as shown in~\cref{sec:gradientbased_system_identification}. This allows further reduction of the discrepancy between the physics-based prior mean and the real-world systems.

\subsection{Integrating modular neural network surrogates with beam dynamics simulations}\label{sec:integrating_modular_neural_network_surrogates_with_beam_dynamics_simulations}

Some beam dynamics such as collective effects require expensive computations to simulate.
This problem has previously been solved using \ac{NN} surrogate models. Data from experiments or high-fidelity simulations can be used to train an \ac{NN} to approximate the real world or a high-fidelity simulation with a high degree of accuracy, while forward passes of \acp{NN} are cheap to compute.
To date, single \ac{NN} surrogates are usually used to model, for example, specific instruments or lattice setups~\cite{kaiser2023machine,edelen2017using,ratner2021recovering}.
As a result, these models have limited versatility and reusability, and novel applications require the design and training of new models, which necessitates further beam time or computation to acquire new training data sets.

\textit{Modular surrogate models} over all parameters of generic accelerator elements can be used as a versatile, reusable, and reconfigurable approach to modelling larger parts of accelerators using \acp{NN}.
This modular approach integrates well with Cheetah, as is shown in \cref{fig:cheetah_application_overview} (d). Modular \ac{NN} surrogates computing expensive physical effects can seamlessly be wrapped as Cheetah elements and combined with other elements using beam dynamics models already implemented in Cheetah or other \ac{NN} surrogates.
Crucially, \acp{NN} are differentiable and commonly implemented in PyTorch. Hence, they integrate well with Cheetah's PyTorch backend and preserve PyTorch's automatic differentiation functionality.

\begin{figure}
    \centering
    \includegraphics{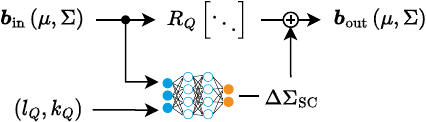}
    \caption{Scheme of the \ac{NN}-enhanced quadrupole module. The incoming \texttt{ParameterBeam} $P$ is multiplied with the magnet's transfer matrix $\transfermap_\quadrupole$ as in classical linear beam dynamics. The \ac{NN} model predicts changes $\spacechargedelta$ in beam parameters due to the space charge effects, resulting in the outgoing beam $\measuredbeam_\text{out}$.}
    \label{fig:nn_enhanced_quadrupole_scheme}
\end{figure}

In this use case example, we demonstrate the implementation of a quadrupole magnet augmented with space charge effects modelled using a \ac{NN} and integrate it as an element in Cheetah.
The straightforward implementation of an \ac{NN}-based modular surrogate model would see the \texttt{track} method in Cheetah implemented as a forward pass of an \ac{NN}

\begin{equation}
    \outgoingbeam = f_\text{SC} \left( \incomingbeam | l_\quadrupole, \quadstrength_\quadrupole \right) \text{,}
\end{equation}

mapping the incoming beam $\incomingbeam$ and quadrupole parameters $\left( l_\quadrupole, \quadstrength_\quadrupole \right)$ to an outgoing beam $\outgoingbeam$.
Fortunately, the effects of space charge in a quadrupole are secondary to the linear beam dynamics, and linear beam dynamics can be modelled easily. To reduce the complexity of the function modelled by the \ac{NN} and reduce the time and data required to train it, the tracking function through the quadrupole element is instead reformulated as

\begin{equation}
    \beam_\text{out} = f_\text{linear} \left( \incomingbeam | l_\quadrupole, \quadstrength_\quadrupole \right) + \spacechargedelta \left( \incomingbeam | l_\quadrupole, \quadstrength_\quadrupole \right) \text{,}
\end{equation}

where $f_\text{linear} \left( \incomingbeam | l_\quadrupole, \quadstrength_\quadrupole \right)$ is the handcrafted computation of the linear beam dynamics through a quadrupole magnet already implemented by Cheetah and $\spacechargedelta \left( \incomingbeam | l_\quadrupole, \quadstrength_\quadrupole \right)$ is the change induced to the outgoing beam by space charge effects.
The \ac{NN} model is used to approximate $\spacechargedelta \left( \incomingbeam | l_\quadrupole, \quadstrength_\quadrupole \right)$. An illustration of this process is provided in \cref{fig:nn_enhanced_quadrupole_scheme}.

Data for training the modular \ac{NN} surrogate model is generated using Ocelot~\cite{agapov2014ocelot} with space charge effects and second-order tracking through a single quadrupole element.
Space charge effects are calculated with a mesh size of $63^3$ and applied at a unit step size of \qty{2}{\centi\meter}.
A total of \num{100000} samples are collected from uniform distributions over length $l_\quadrupole$ and strength $\quadstrength_\quadrupole$ of the quadrupole, as well as log-uniform distributions over a subset of the incoming beam parameters

\begin{equation}
    \nninput \coloneqq \left( \sigma_x, \sigma_{x'}, \sigma_y, \sigma_{y'}, \sigma_\tau, \sigma_\delta, E, q \right) \subset \incomingbeam \text{,}
\end{equation}

where $E$ is the beam energy and $q$ is the total charge of the beam. The ranges over which these are sampled for data generation are shown in \cref{tab:data_ranges}. A log-uniform distribution was chosen for the beam input parameters because their order of magnitude is more relevant in space charge computations.

\begin{table}
    \centering
    \caption{Input parameter ranges for data set generation}
    \label{tab:data_ranges}
    \begin{tabular}{lc}
        \toprule
        \textbf{Input parameter} & \textbf{Range}                                                            \\
        \midrule
        $\sigma_x$               & $\left[ \qty{10}{\micro\meter}, \qty{1}{\milli\meter} \right]$            \\
        $\sigma_{x'}$            & $\left[ \qty{10}{\micro\radian}, \qty{1}{\milli\radian} \right]$          \\
        $\sigma_y$               & $\left[ \qty{10}{\micro\meter}, \qty{1}{\milli\meter} \right]$            \\
        $\sigma_{y'}$            & $\left[ \qty{10}{\micro\radian}, \qty{1}{\milli\radian} \right]$          \\
        $\sigma_{\tau}$          & $\left[ \qty{300}{\nano\meter}, \qty{300}{\micro\meter} \right]$          \\
        $\sigma_\delta$          & $\left[ \num{e-5}, \num{e-3} \right]$                                     \\
        $q$                      & $\left[ \qty{1}{\pico\coulomb}, \qty{5}{\nano\coulomb} \right]$           \\
        $E$                      & $\left[ \qty{1}{\mega\electronvolt}, \qty{1}{\giga\electronvolt} \right]$ \\
        \bottomrule
    \end{tabular}
\end{table}

The neural network model takes $\nninput$ as input and outputs
\begin{equation}
    \nnoutput \coloneqq \left( \Delta\sigma_x, \Delta\sigma_{x'}, \Delta\sigma_y, \Delta\sigma_{y'}, \Delta\sigma_\tau,\Delta\sigma_\delta \right)
\end{equation}
the changes to the beam parameters resulting from space charge effects when compared to linear beam dynamics, such that
\begin{equation}
    \spacechargedelta \left( \incomingbeam | l_\quadrupole, \quadstrength_\quadrupole \right) \approx \nnoutput = f_\text{NN} \left( \nninput \right) \text{.}
\end{equation}

We choose an \ac{MLP} architecture for the \ac{NN} and the Adam~\cite{kingma2014adam} gradient descent algorithm for adjusting the parameters of the model. Early stopping with a \textit{patience} (number of steps with no improvement before the training is terminated) of \num{10} was used. The data set is split 60/20/20 into training, validation, and test sets. The logarithm is taken of all beam parameter inputs before they are input into the \ac{NN} model. All inputs and outputs are scaled to fit a unit-normal distribution with scaling on the beam parameter inputs performed after taking the logarithm.
Hyperparameters were tuned over a total of \num{303} trainings using Bayesian optimisation, with \textit{PyTorch Lightning}~\cite{falcon2019pytorch} used to implement the training and \textit{Weights \& Biases}~\cite{biewald2020experiment} for experiment tracking.
The best-observed hyperparameters used for the final model are listed in \cref{tab:hyperparameters}.

\begin{table}
    \centering
    \caption{\Ac{NN} training hyperparameters}
    \label{tab:hyperparameters}
    \begin{tabular}{lc}
        \toprule
        \textbf{Hyperparameter}    & \textbf{Value}               \\
        \midrule
        Batch size                 & \num{32}                     \\
        Hidden activation          & Sigmoid                      \\
        Hidden layer width         & \num{256}                    \\
        Learning rate              & \num{2e-5}                   \\
        Number of epochs           & \num{959} (max. \num{10000}) \\
        Number of hidden layers    & \num{4}                      \\
        Gradient descent algorithm & Adam                         \\
        \bottomrule
    \end{tabular}
\end{table}

\begin{figure*}
    \centering
    \includegraphics{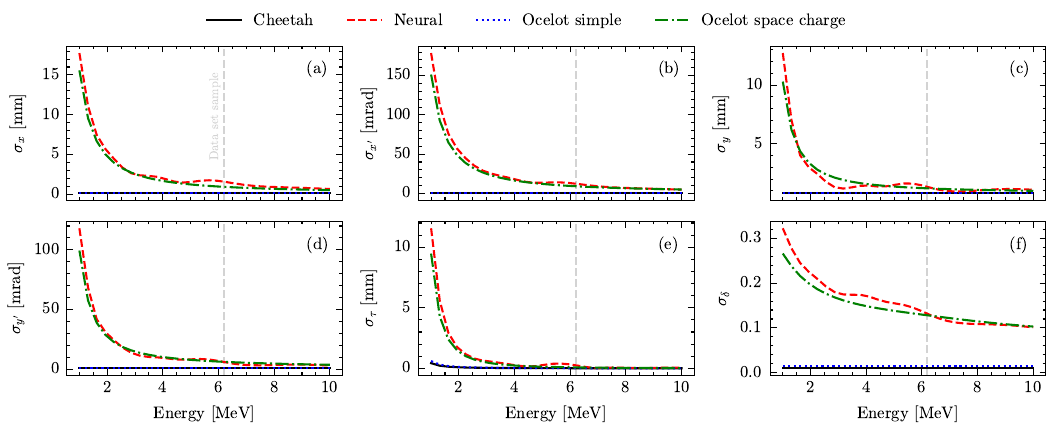}
    \caption{Outgoing beam parameters for the same incoming beam over different energies tracked with the default linear quadrupole in Cheetah (black), the \ac{NN}-augment space charge quadrupole (red), Ocelot without space charge calculations (blue), and Ocelot with space charge calculations (green).}
    \label{fig:energy_scan_example}
\end{figure*}

\Cref{fig:energy_scan_example} shows the beam sizes, beam divergence, bunch length, and energy spread over different beam energies as computed by the \ac{NN}-augmented quadrupole implemented in Cheetah, and compares them to Cheetah's default linear beam dynamics tracking and Ocelot's space charge simulation. All other parameters of the incoming beam and the quadrupole parameters are fixed. We observe that the \ac{NN}-augmented quadrupole implementation correctly infers larger effects of space charge at low energies when compared to linear beam dynamics simulations without space charge effects. Congruently, the beam parameters computed using the \ac{NN}-augmented quadrupole in Cheetah closely match those computed using Ocelot with space charge effects, indicating that our simulations accurately capture space charge effects.

We benchmark the computational speed of the \ac{NN}-augmented space charge quadrupole element against Ocelot's space charge simulation. For this evaluation, the incoming beam and quadrupole settings considered for the experiments in \cref{fig:energy_scan_example} are used with an energy of \qty{6.2}{\mega\electronvolt}. Ocelot is configured with the same space charge simulation setup as was used to generate the training data. We find that Ocelot takes an average of \qty{1.36}{\second} to perform space charge tracking through a single quadrupole, while the same simulation is performed by the \ac{NN}-augmented Cheetah element in an average of \qty{370}{\micro\second} -- a reduction in compute time of more than \num{3} orders of magnitude. These benchmarks were run on the same Apple M1 Pro SOC's \ac{CPU} considered in \cref{sec:speed}. Unlike Ocelot's space charge implementation, Cheetah can take advantage of hardware acceleration on GPU, which is expected to result in a further reduction of compute times.

Moreover, because an \ac{NN} is used for the modular surrogate modelling, the computation remains fully differentiable, in effect providing differentiable space charge simulations that can seamlessly be integrated with other beam dynamics simulations.

This example serves as a proof of concept for Cheetah's ability to provide a platform for modular surrogate modelling. In fact, to the best of our knowledge, this is the first instance of modular \ac{NN}-based surrogate modelling in particle accelerator simulations.
As such, the integration of modular \ac{NN} surrogate models in Cheetah enables us to build data-driven, high-speed, high-fidelity models of beam dynamics that would otherwise require computationally expensive models. Moreover, Cheetah can also be integrated with modular \ac{NN} surrogates trained on real-world data, allowing for mitigation of model mismatches.
In future work, we hope to add more \ac{NN}-augmented and fully \ac{NN}-based elements to Cheetah. While the presented example applies to parametrically defined beams, it can easily be extended to beams defined as particle clouds by employing \ac{NN} architectures such as \textit{PointNet}~\cite{qi2016pointnet}, which is intended as a future extension of Cheetah.

\section{Conclusions}

In this work, we introduced Cheetah, a Python package providing high-speed differentiable beam dynamics simulations for machine learning applications. Cheetah is easy to use, provides an extensible platform for future differentiable models and integrates well with the \ac{ML} ecosystem in Python.
Moreover, we demonstrated Cheetah's capabilities using five example applications.
We illustrated its speed training a \ac{NN} policy to perform transverse beam tuning while achieving zero-shot transfer to the real world.
Further, we showed how automatic differentiation in Cheetah can be used for gradient-based beamline tuning as well as gradient-based system identification.
Cheetah's usefulness as a differentiable prior for Bayesian optimisation was also shown while optimising beam focusing through a FODO cell.
Lastly, we presented an example of how Cheetah can easily be extended by training a modular \ac{NN} space charge model to predict how space charge affects the beam when tracked through a single quadrupole magnet.

Cheetah will see continued extension as a tool for our work.
We further hope that in the future, members of the community will collaborate in extending Cheetah, for example with already developed differentiable models of processes in particle accelerators. Such collaboration and integration will help make these tools more accessible to the community.
Further extensions of Cheetah planned from our side include the integration of additional modular surrogate models, in particular for single particle tracking based on PointNet~\cite{qi2016pointnet}, and batched parallel execution of simulations to increase speeds further.
In addition, experiments using JAX~\cite{jax2018github} as a backend for Cheetah might aid in attaining even faster simulation speeds as was seen by switching the backend of Stable Baselines3 to JAX in SBX~\cite{raffin2021stable}.

\section*{Code availability}

The source code of Cheetah is hosted at \url{https://github.com/desy-ml/cheetah}.
The code for the presented use case examples is available from \url{https://github.com/desy-ml/cheetah-demos}.
More extensive code regarding the \ac{RL} example may be found at \url{https://github.com/desy-ml/rl-vs-bo}.

\begin{acknowledgments}
    This work has in part been funded by the IVF project InternLabs-0011 (HIR3X) and the Initiative and Networking Fund by the Helmholtz Association (Autonomous Accelerator, ZT-I-PF-5-6).
    The authors acknowledge support from DESY (Hamburg, Germany) and KIT (Karlsruhe, Germany), members of the Helmholtz Association HGF, as well as support through the \textit{Maxwell} computational resources operated at DESY and the \textit{bwHPC} at SCC, KIT. The authors thank Oliver Stein for his contribution to JOSS, from which Cheetah was derived. Furthermore, the authors thank Felix Theilen for the software maintenance work he performed on Cheetah. Finally, the authors thank Frank Mayet for providing help with the parallel version of ASTRA.
\end{acknowledgments}

\section*{Author contributions}

J.K. originally conceived and developed Cheetah as a fast simulation code.
Further development of Cheetah as a differentiable simulation code was done by J.K. and C.X.
J.K. undertook example studies of Cheetah on reinforcement learning, gradient-based tuning, gradient-based system identification, and modular neural network surrogate modelling.
C.X. performed example studies on using Cheetah as a prior for Bayesian optimisation and contributed to the other examples studies.
J.K. wrote the manuscript.
C.X. contributed sections to the manuscript and provided substantial edits.
A.E. and A.S.G. secured funding.
All authors read and edited the manuscript.


\bibliography{bibliography} 

\begin{acronym}
    \acro{ANN}{artificial neural networks}
    \acro{BPM}{beam position monitor}
    \acro{BO}{Bayesian optimisation}
    \acro{CI}{continuous integration}
    \acro{CPU}{central processing unit}
    \acro{CUDA}{Compute Unified Device Architecture}
    \acro{EI}{expected improvement}
    \acro{EA}{Experimental Area}
    \acro{ES}{extremum seeking}
    \acro{FDF}{focus-defocus-focus}
    \acro{GAN}{generative adversarial network}
    \acro{GP}{Gaussian process}
    \acro{GPGPU}{general purpose GPU}
    \acro{GPU}{graphics processing unit}
    \acro{HPC}{high-performance computing}
    \acro{MAE}{mean absolute error}
    \acro{ML}{machine learning}
    \acro{MLP}{multilayer perceptron}
    \acro{MPS}{Metal Performance Shaders}
    \acro{MSE}{mean squared error}
    \acro{NN}{neural network}
    \acro{PPO}{Proximal Policy Optimisation}
    \acro{ReLU}{rectified linear unit}
    \acro{RMSE}{root mean squared error}
    \acro{RL}{reinforcement learning}
    \acro{RLO}{reinforcement learning-trained optimisation}
    \acro{SGD}{stochastic gradient descent}
    \acro{TD3}{Twin Delayed DDPG}
    \acro{UCB}{upper confidence bound}
\end{acronym}

\end{document}